\pdfoutput=1

\documentclass[a4paper,10pt,twocolumn]{article}
\usepackage[T1]{fontenc}
\usepackage{lmodern}
\usepackage{lipsum}
\usepackage{caption}
\usepackage{authblk}
\usepackage[top=2cm, bottom=2cm, left=2cm, right=2cm]{geometry}
\usepackage{fancyhdr}
\usepackage{url}
\usepackage{graphicx}
\usepackage{amssymb, gensymb}
\usepackage{amsmath}
\usepackage[authormarkuptext=id]{changes}
\definechangesauthor[name={Drosos}, color=red]{DK}
\usepackage{soul}
\usepackage{tikz}
\usepackage{pgfplots}
\pgfplotsset{compat=1.8}
\usetikzlibrary{patterns}
\usepackage{pgfplotstable, booktabs}  
\definecolor{mycolor1}{RGB}{130,220,202}
\definecolor{mycolor2}{RGB}{79,122,142}  
\definecolor{mycolor3}{RGB}{170,35,3}
\definecolor{mycolor4}{RGB}{207,170,114}
\definecolor{mycolor5}{RGB}{80,135,63}
\definecolor{mycolor6}{RGB}{255,140,190}

\begin{document}

  \title{\sf Micromagnetic Simulations Study of Skyrmions in Magnetic FePt  Nanoelements} 
 \author[1]{Leonidas N. Gergidis\thanks{lgergidi@uoi.gr;\; lgergidis@gmail.com} } 
 \author[1]{Vasileios D. Stavrou}
 \author[2]{Drosos Kourounis} 
 \author[1]{Ioannis Panagiotopoulos}
\affil[1]{Department of Materials Science and Engineering, University of Ioannina, 45110 Ioannina, Greece}
\affil[2]{NEPLAN AG, CH-8700  K\"usnacht (ZH), Switzerland}

\maketitle
\thispagestyle{empty}
\begin{abstract}
  The magnetization reversal in $330 nm$ triangular prismatic magnetic 
  nanoelements with variable magnetocrystalline anisotropy similar to that of  partially chemically ordered FePt is studied using 
  micromagnetic simulations employing Finite Element discretizations.
  Several magnetic properties including the 
  evaluation of the magnetic 
  skyrmion number $S$ are computed in order to 
  characterize magnetic 
  configurations exhibiting vortex-like formations. 
  Magnetic vortices and skyrmions are revealed in different systems generated by the variation of the magnitude and relative orientation of
  the magnetocrystalline anisotropy direction, with respect to
  the normal to the triangular prism base. Micromagnetic configurations with skyrmion number greater than one   
  have  been  detected  for  the  case  where  magnetocrystalline anisotropy was normal to nanoelement's base. For particular magnetocrystalline anisotropy values  three distinct skyrmions 
  are formed  and persist for a range 
  of external fields. The simulation-based
  calculations of the skyrmion number $S$ revealed that skyrmions   can be created for magnetic nanoparticle systems 
  lacking of chiral interactions such as Dzyaloshinsky-Moriya, but by only varying the magnetocrystalline anisotropy.
\end{abstract}
%
\date{}

\section{Introduction}
\label{sec:intro}
Magnetic nanoparticles and nanostructures find 
numerous applications in a wide variety
of scientific fields including information signal processes, spin
devices, high density storage media, magnetic sensors, medicine and biology \cite{Parkin:2008,Novosad:2010,Sampaio:2013,Hovorka:2017,Iwasaki:2014}.
Nowadays progress in nanoscale material growth have allowed 
 the synthesis and fabrication of nanoparticles in a wide
range of shapes and sizes \cite{Sellmyer:2006,Wang:2012,Jeong:2001}. 
Their magnetic response is of paramount importance and could be associated with geometrical and materials factors; thus, a lot of effort has been devoted to experimental, simulation and theoretical studies. In particular, the process of magnetization reversal in
magnetic nanoparticles could be exploited, engineered for technological applications but necessitates the knowledge and control over the formation of rather complex 
micromagnetic structures, such as multiple domain walls, vortices, skyrmions-antiskyrmions and merons \cite{Stavrou:2016,YuPnas:2012,Tan:2016,Xia:2017,Xia:2018,Tomasello:2015,Muller:2017,Jaafar:2010}.

Magnetic skyrmions are vortex-like magnetization
configurations. They have been predicted theoretically by Bogdanov et al. \cite{Bogdanov:1994},\cite{Bogdanov:1995} long before their
experimental detection and discovery \cite{Ding:2014,Li:2014}.  In recent years magnetic skyrmions have attracted a lot of theoretical
\cite{Heinze:2011,Nagaosa:2016,Nagaosa:2017,Stier:2017,Buttner:2018,Komineas:1996}, simulation \cite{FangohrNov:2015,Jin:2017,Stosic:2017,Fangohr:2018}, and experimental
\cite{Heinze:2011,Romming:2015,Jiang:2015,Hanneken:2015,Moutafis:2016} attention 
due to their thermodynamic and topological stability, their small
size, and their inherent property of easy movement and repositioning under the application of low or even tiny in-plane electric
currents. They seem promising for use 
in next generation spintronic 
devices \cite{Romming:2013,Zhang:2015} as information carriers, giving the credentials of ultra dense low-cost power 
storage and the capability to perform logical operations \cite{He:2017}. The next generation magnetic memory devices would rely on the efficient
creation and control of magnetic textures, 
such as magnetic skyrmions
using rather tiny electric currents
\cite{Sampaio:2013,Nakatani:2016,Fert:2017,Sitte:2017}.
The aforementioned topological stability is related
to the confined skyrmion magnetic configuration, which is 
predicted to be stable because the 
individual atomic spins, oriented opposite to those of the 
surrounding thin-film cannot perform flipping motion.
Spins hindered to align themselves with the rest of atoms 
in confined geometry  without overcoming an energy barrier. 
The origin of the energy barrier is attributed to the
``topological protection''. 
The thermodynamic stability of skyrmions 
is considerably strong and can be attributed to the particular magnetic configuration which can be characterized
by a total topological charge described by skyrmion number $S$
\cite{Nagaosa:2016,Nagaosa:2017}. This skyrmion number $S$ is an
integer and to this point is being considered having quantized values
that cannot be changed continuously. The skyrmion number $S$ is defined as
\begin{equation}\label{eq:skyrmion}
S  = \frac{1}{4\pi} \int_A q_{Sk} \; dA 
\end{equation}
where $q_{Sk}$  is given by the following relation
\begin{equation}\label{eq:chargedensity}
q_{Sk} =\mathbf{m} \cdot (\frac{\partial
	\mathbf{m}}{\partial x} \times \frac{\partial \mathbf{m}}{\partial
	y}).
\end{equation}
The quantity $\bf{m}$ is
the unit vector of the local magnetization defined as
$\mathbf{m}=\mathbf{M}/M_s$ with $\mathbf{M}$ being the magnetization
and $M_s$ the saturation magnetization. The skyrmion number $S$ is a physical and topological quantity 
that  measures how many times $\mathbf{m}$ wraps the unit sphere 
\cite{Yoo:2014}. The
integrated quantity describes the topological 
density $q_{Sk}$ and has units of $\mathrm{nm^{-2}}$, which are implied throughout the manuscript.
In many instances the integrated quantity is also referred as
``topological charge''. 
Surface $A$ is the surface
domain of integration and corresponds to the upper or the lower
triangular bounding surface of the FePt nanoelement under
investigation.

The magnetization reversal in 330 nm triangular prism magnetic
nanoelements with variable magnetocrystalline 
anisotropy (as that of
partially chemically ordered FePt) has been studied using Finite
Elements micromagnetic simulations in  \cite{Stavrou:2016}. 
The simulation results showed that a wealth of 
reversal mechanisms is possible
sensitively depending on the uniaxial magnetocrystalline anisotropy values and directions; the latter may explain the different Magnetic Force Microscopy patterns obtained in such magnetic systems.
In addition, the micromagnetic simulations revealed that interesting vortex-like formations can be produced and stabilized in large field ranges and in sizes that can be tuned by the magnetocrystalline anisotropy (MCA) of the material.  The aforementioned spontaneous  ground states of skyrmion-like
configurations were obtained and in other magnetic systems 
without the implication of chiral
interactions such as Dzyaloshinsky-Moriya (DMI) \cite{Xia:2018,Stier:2017,Dai:2013,Zhou:2015,Sapozhnikov:2015,Guslienko:2015,Zhang:2016}. 

In the present  work  a quantitative description 
is given for the vortex-skyrmionic   
configurations by calculating the skyrmion number
at $0\degree$K obtained for FePt triangular 
magnetic nanoislands having variable 
magnetocrystalline anisotropy.
The aforementioned calculation can reveal information not readily
recognizable by simple visual-inspection of the micromagnetic
configurations. Furthermore, the skyrmion 
number as a function of the
applied field along a hysteresis curve can give quantitative
information of the reversal mechanisms and energy barriers involved. In what follows we present examples on the numerical calculation of skyrmion number as means of characterizing magnetic configurations and
reversal in nanoelements including thin film asymmetric triangular nanoislands.
 
\section{Micromagnetic modeling}
\label{sec:model}
\subsection{FEM solution of Landau-Lifshitz-Gilbert (LLG) equation}
The rate of change of the 
dynamical magnetization field $\mathbf{M}$ is
governed by a nonlinear equation of motion, the Landau-Lifshitz-Gilbert (LLG) equation
\begin{equation}
\frac{d\mathbf{M}}{dt}=\frac{\gamma}{1+\alpha^2}(\mathbf{M} \times \mathbf{H}_{eff})
-\frac{\alpha \gamma}{(1+\alpha^2)|\mathbf{M}|} \mathbf{M} \times (\mathbf{M}\times \mathbf{H}_{eff}).
\end{equation}
In the aforementioned LLG equation $\alpha>0$ is a phenomenological
dimensionless damping constant that depends on the material and
$\gamma$ is the electron gyromagnetic ratio. The effective field that governs the dynamical behavior of the system has contributions 
from various effects that are of very 
different nature and can be expressed as 
$\mathbf{H}_{eff}=\mathbf{H}_{ext}+\mathbf{H}_{exch}+
\mathbf{H}_{anis}+\mathbf{H}_{demag}.$
Respectively, these field
contributions are the external magnetic field $\mathbf{H}_{ext}$,
the exchange field $\mathbf{H}_{exch}$, the anisotropy
field $\mathbf{H}_{anis}$ and the demagnetizing field $\mathbf{H}_{demag}$.

\begin{figure}[!b]
\centering
\includegraphics[totalheight=1.6in,angle=0]{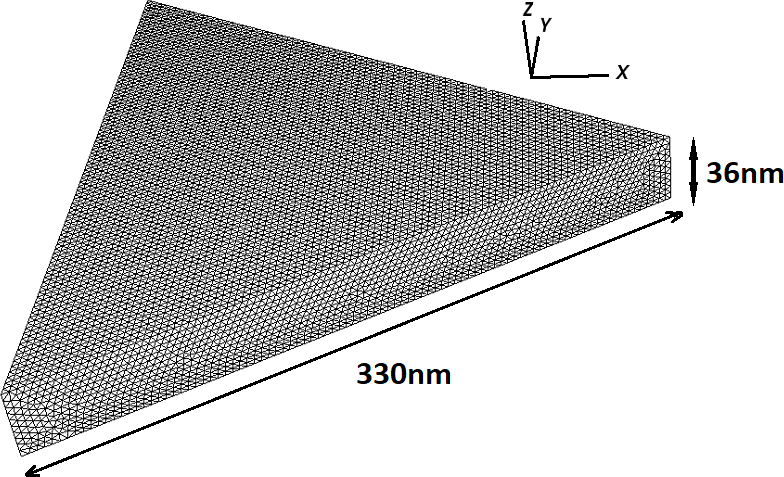} 
	\caption{Model geometry and the generated mesh.}
	\label{fig:geometrymesh}
\end{figure}

For the solution of the LLG equation We have 
performed micromagnetic finite element calculations using 
the Nmag software \cite{Nmag:2007}.
The dimensionless damping constant $\alpha$ was set to 1  
in order to achieve fast damping and
reach convergence quickly as we are interested in static magnetization configurations. The default convergence 
criterion for each applied external 
field step $\mathbf{H}_{ext}$
was that the magnetization should move slower than 1 degree per
nanosecond, globally or on average for all spins.
The sample was described as a triangular prism with equilateral
triangle base of 330 nm and 36 nm in height and is shown in  
\textbf{Fig. \ref{fig:geometrymesh}}. 
The 36 nm thickness of the
film matches that reported in~\cite{Markou:2013} while
Okamoto in \cite{Okamoto:2002} suggests 24 nm, which 
is comparable and not 
expected to lead to qualitatively different behavior. The following frame of reference axes assignment convention was used: x along the triangle height, y along the side perpendicular to x, and z perpendicular to the film plane.
The considered finite element mesh for the triangular film under study was generated using the automatic three dimensional (3D) tetrahedral
mesh generator Netgen \cite{NETGEN:2007}.  
We have used a 3.4 nm
maximum distance between nodes which is 
lower than the value of the
exchange length $l_{ex}=\sqrt{\frac{2A}{\mu_0M_s^2}} \approx 3.5 $ nm. This resulted in 488874 volume elements (94800 points) per triangular magnetic island.  The material parameters 
were chosen similar to those
typical for bulk FePt with a saturation 
polarization of ${\mu_0
  M_s=1.43T}$ ($M_S$=1.138MA/m), and an exchange constant
$A_{exch}$=11pJ/m which has been found to 
be independent of the degree
of ordering Okamoto \cite{Okamoto:2002}. 
The magnetocrystalline
anisotropy (MCA) constant $Ku$ was varied 
between $Ku\mathrm{=100kJ/m^3}$ and 
$\mathrm{500kJ/m^3}$, as for $Ku$ 
exceeding this value the reversal mode was
simply homogeneous rotation. The demagnetization 
factor $\cal{N}$ has been estimated to ${\cal{N}}=0.71$ by 
the saturation field perpendicular to the 
triangle for the case of $Ku=\mathrm{0kJ/m^3}$. We must note 
that for the range of these MCA values the 
easy axis remains in-plane as the 
shape anisotropy contribution $-\frac{1}{2}{\cal N}\mu_0 M_s^2=\mathrm{-580kJ/m^3}$ exceeds $Ku$ for all the cases presented here. Four different directions of the
magneto-crystalline anisotropy were 
tested along $x,y,z$ as well as
along the $[111]$ direction. The latter is of interest as in many instances FePt and CoPt films tend to grow with their $<111>$ crystallographic directions along the film normal resulting an angle of $54.7\degree$ to the film normal \cite{Alexandrakis:2009}.

The total duration of the micromagnetic simulation was ranging for 1 to 10 days on Intel i7 4770K 
depending on the relative orientation of the applied
magnetic field with respect to magnetic 
anisotropy. The magnetization
curves for every production run were 
investigated by applying external 
magnetic fields $H_{ext}$ with fixed orientation 
running parallel to z-direction (the normal to the 
triangular base). The range values of  $H_{ext}$ were 
 +1000kA/m (maximum) and -1000kA/m (minimum) 
 introducing an external magnetic field
step $\delta H_{ext}$=4kA/m \cite{Stavrou:2016}.
\subsection{Skyrmion number computation}
The calculation of skyrmion number $S$ necessitates the
knowledge of the computed normalized magnetization 
vector $\mathbf{m}$ obtained from 
the solution of LLG equations. 
Once the finite element approximation of the normalized magnetization $\mathbf{m}_h$ has been computed, the skyrmion number $S_h$, 
following \eqref{eq:skyrmion}, is approximated by
\begin{equation}
    \label{eq:DiscreteSkyrmion}
    S_h 
     = \frac{1}{4\pi} \sum_{e=1}^{N_e} \mathbf{m}_h^e \cdot \left( \frac{\partial \mathbf{m}_h^e}{\partial x}
    \times \frac{\partial \mathbf{m}_h^e}{\partial y} \right ) |A_e|.
    %
\end{equation}
where $m_h,S_h$ are the discrete representations of $m,S$ respectively with $e$ denoting the element and $N_e$ the total number of elements used for the surface 
domain discretization. It should be
noted that throughout the manuscript the symbol $S$ 
instead of $S_h$ will be used for the computed value of the 
skyrmion number. 
The integration takes place over 
the top or bottom surface boundary of the prism $A$.
Since we are using tetrahedral P1 elements for the discretization of the prism volume,
the top (or bottom) surface boundary is comprised of triangles with outwards pointing normal parallel 
to the $\hat{\mathbf{z}}$ unit vector. 
Magnetizations are extracted for the top  
or bottom surface of the magnetic triangular 
prismatic nanoelement. These particular  magnetizations 
located on the surface elements of the two bases are used for 
the actual computations of $S$ and of the relative topological
quantities.  
It is possible for a magnetic configuration to include more than
one skyrmion. Inevitably, the total skyrmion number would be the
algebraic sum of its  individual skyrmionic
configurations. It follows that a structure that locally includes several skyrmions of different polarity 
or chirality may yield a zero
total skyrmion number. The physical
significance of this situation 
may be attributed to the fact that
structures with opposing $S$ may be easier to mutually
annihilate. Therefore, it is of interest to monitor following 
\cite{FangohrNov:2015} the integral of
the absolute value of the topological density symbolized with
$S_{abs}$ as it describes the existence of topological entities that are masked and washed out when 
$S$ is calculated through the integral
over all surface domain $A$. The quantity $S_{abs}$ is defined by the relation
\begin{equation}\label{eq:absskyrmion}
S_{abs}= \frac{1}{4\pi} \int_A \left | \mathbf{m} \cdot \left (\frac{\partial \mathbf{m}}{\partial x} \times  \frac{\partial \mathbf{m}}{\partial y} \right ) \right | dA
\end{equation}
The scalar quantity $S_{abs}$ is injective and provides the
necessary distinctness for different skyrmionic 
states \cite{FangohrNov:2015}.
The stabilization of such magnetic skyrmions is usually 
linked to the existence of some kind of anisotropic Dzyaloshinskii-Moriya interaction (DMI). Its discrete estimation follows  similarly to \eqref{eq:DiscreteSkyrmion}.
It is interesting the calculation of skyrmion numbers for micromagnetic configurations in nanoelements since
it can reveal information not readily recognizable by simple visual inspection of the micromagnetic 
configurations. Furthermore, the skyrmion
number $S$ as a function of the applied field $\mathbf{H}_{ext}$ along an hysteresis curve can provide quantitative information of the reversal mechanisms and energy barriers involved in the process.
\section{Results}
\label{sec:Results}
\subsection{In-plane MCA}
\begin{figure}[!b]
\centering
	\includegraphics[totalheight=1.6in,angle=0]{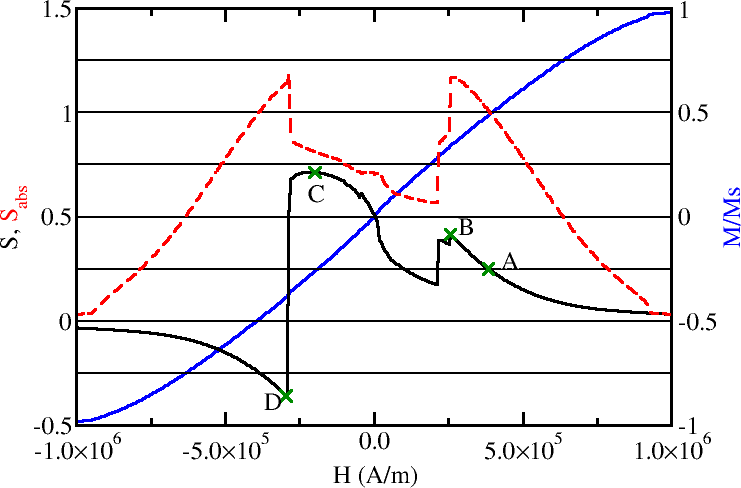}
	\caption{Normalized magnetization ($M/M_s$) and skyrmion numbers $S,S_{abs}$ during   the magnetization reversal process for in plane MCA (along x-direction) with $Ku\mathrm{=100kJ/m^3}$ and $\mathbf{H}_{ext}//[001]$.}
	\label{fig:HzKxKu100skyrmionloop}
\end{figure}

\begin{figure}[!t]
\centering
\includegraphics[totalheight=3.0in,angle=0]{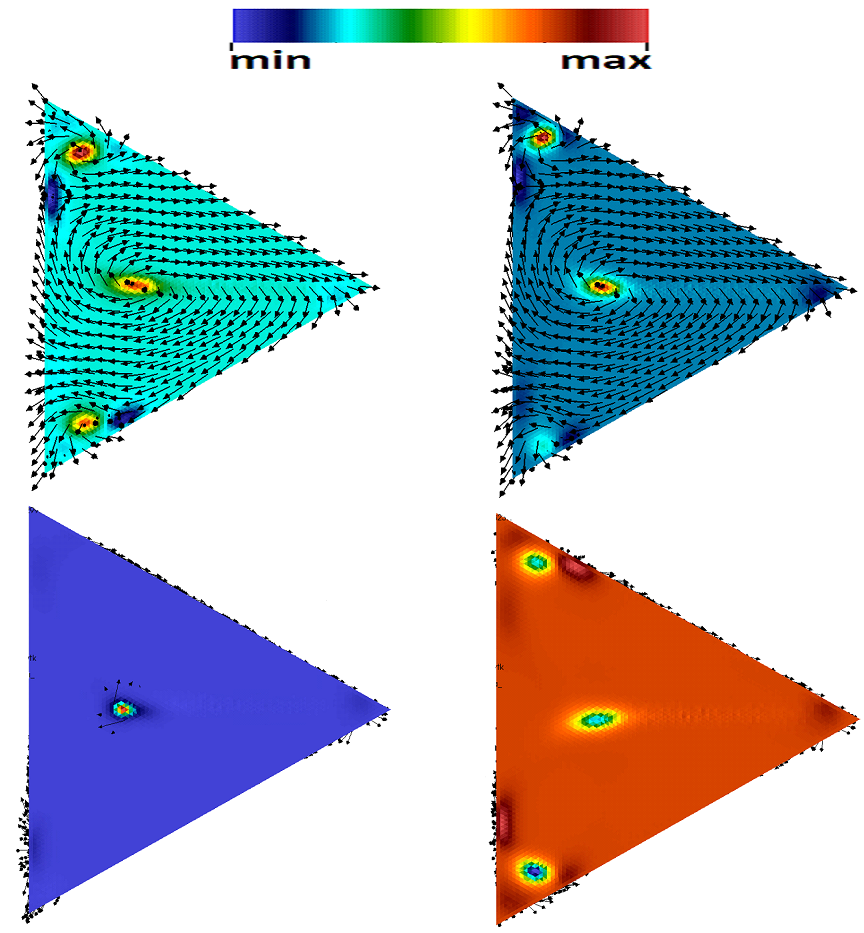} 
	\caption{Top view of $q_{Sk}$  at $Ku\mathrm{=100kJ/m^3}$ for MCA being parallel to x-direction and $\mathbf{H}_{ext}//[001]$. From left to right and from top to bottom different values of  ${H}_{ext}=384 ,256,-200,-296 \; (\times \mathrm{10^3 A/m})$ are represented referring to points of A, B, C, D of  \textbf{Fig. \ref{fig:HzKxKu100skyrmionloop}}. 
	Local magnetization vectors are shown with arrows superimposed with $q_{Sk}$. The actual value ranges of $q_{Sk}/nm^{-2}$ are represented in color bars with maximum and minimum values respectively: left top (A  $max:0.00394\; -\; min:-0.00174$) , right top 
	(B  $max:0.00693\; -\; min:-0.00203$ ), left down (C  $max:0.0272\; -\; min:-0.00037$), right down	(D  $max:0.00185\; -\; min:-0.00057$).}
	\label{fig:HzKxSsnapshots}
\end{figure}

 In the first system studied the 
 magnetocrystalline anisotropy (MCA) lies within 
the plane of the triangular nanoelement parallel to
$x$-direction ($\mathbf{K}//[100]$)
 with the external field $\mathbf{H}_{ext}$ applied along
 the width of the nanoelement which is parallel 
 to $z$-direction ($\mathbf{H}_{ext}//[001]$). 
 The $\mathbf{H}_{ext}$ direction 
is being fixed parallel to $z$-direction 
throughout this work for all systems studied.  
We have calculated the
half-hysteresis loop (descending branch of the loop) for the triangular prismatic 
nanoelements for different MCA
values. The $Ku\mathrm{=100kJ/m^3}$ is 
presented in \textbf{Fig. \ref{fig:HzKxKu100skyrmionloop}} along 
with $q_{Sk}$. Some micromagnetic configurations 
being formed along the path of the reversal are shown in 
\textbf{Fig. \ref{fig:HzKxSsnapshots}}. We present 
results only for a declining external field $H_{ext}$ since the full hysteresis diagram does 
not contribute any additional
information regarding the magnetization reversal process \cite{Stavrou:2016}. Presenting the full hysteresis diagram would inevitably doubled the required computational effort.
%

\begin{figure}[!t]
	\centering
		\includegraphics[totalheight=1.6in,angle=0]{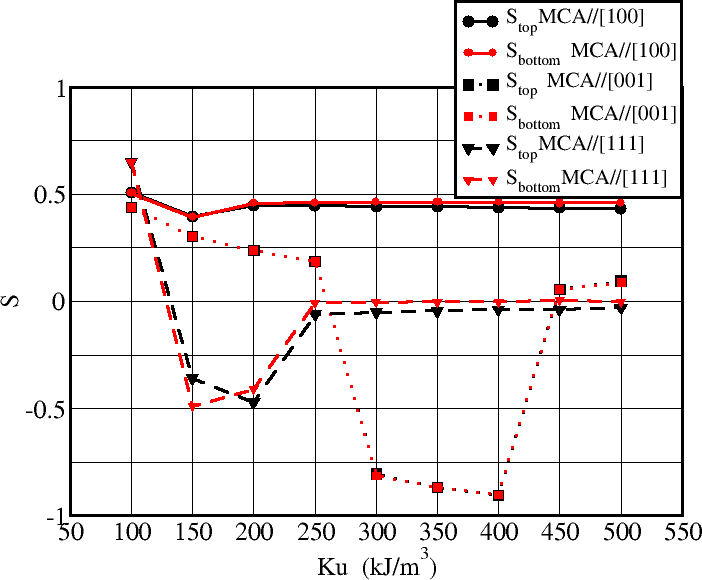}
		\caption{$S$ as a function of  $K_u$ for MCA parallel to [100], [001] and [111] directions with $\mathbf{H}_{ext}//[001]$. $S$ is being calculated  on the top and bottom surface of the triangular prismatic nanoelement for $H_{ext}=0$A/m.}
		\label{fig:HzKxtopbottom}
\end{figure}

Magnetization reversal depicted in 
\textbf{Fig. \ref{fig:HzKxKu100skyrmionloop}} starts 
when the normalized magnetization
decreases from the the saturation value  $M/M_s=1$,  
and passing through the nucleation field it reaches to 
$M/M_s=0$. Then after passing the annihilation 
field, it attains finally the value
$M/M_s=-1$ indicative that all spins have reversed their magnetization vector orientation.  During this reversal process the values of $S$ and $S_{abs}$ are computed 
in order to provide a quantitative
description of the skyrmion-like localized configurations in
conjunction with the qualitative actual visualization of the normalized magnetization vector 
$\textbf{m}$ of the individual spins.  
Values of $S$ have been calculated for 
the top and bottom surface of the
triangular prismatic nanoelement for various $Ku$ 
values shown in \textbf{Fig. \ref{fig:HzKxtopbottom}}
for all magnetic systems 
in the present investigation.  The values of $S$ have the
same quantitative and qualitative behavior 
on top and bottom surface
of the nanoelement ensuring the consistency of the 
numerical calculations. 
Throughout the manuscript the reported values of skyrmion numbers
$S, S_{abs}$ refer to the computed values on the top surface of the nanoelement.

For the MCA value $Ku\mathrm{=100kJ/m^3}$ represented in \textbf{Fig.
\ref{fig:HzKxKu100skyrmionloop}} the process of the
reversal of spins gives birth to vortex like formations. 
The value of $S$ emerges from zero and 
gradually increases; attaining value 0.5, characteristic of magnetic
vortex for external magnetic field  
$H_{ext}\mathrm{=4.0\times 10^3 A/m}$, it reaches the maximum value of
$S \approx 0.75$ that can be considered as 
an incomplete skyrmion 
for field value close to $H_{ext}\mathrm{=-2.1\times
10^5 A/m}$. Note that similar non-integer
values of skyrmion number have been reported in confined
helimagnetic nanostructures \cite{FangohrNov:2015} and in
thin confined polygonal nanostructures \cite{Fangohr:2018}.  It is anticipated that in the present finite magnetic system the skyrmion number can be non-integer due to the restricted area of integration 
$A$ in \textbf{Eq. \ref{eq:skyrmion}} and \textbf{Eq. \ref{eq:absskyrmion}} 
and the essential contribution of magnetostatic energy.  The value of  $S=0.5$ describes a vortex like micromagnetic configuration. 

Around the external field value of
$H_{ext}\mathrm{=9.2\times 10^5 A/m}$ 
a small step can be detected in the
variation of $M/Ms$ reflected also in $S$ and magnified in the
injective property $S_{abs}$ indicating the origin of magnetization reversal through vortex or skyrmion type mechanisms.  In addition,
around $H_{ext}=\mathrm{2.6\times 10^5 A/m}$ 
a jump discontinuity is evident
both for $S,S_{abs}$ not captured by the magnetization curve
$M/M_s$. A second jump discontinuity on $S$ is evident around
$H_{ext}\mathrm{=-2.7\times 10^5 A/m}$ 
causing the change of $S$ value
approximately from 0.7 to - 0.4 (relative variation 157\%).  
The aforementioned variation is followed 
by a change of the sign of $S$
characteristic for a change in the polarity of the vortex type
micromagnetic configuration. Further decrease of
the external magnetic field causes the 
continuous decay of $S$ and its
annihilation following the final stage of the reversal process. 
The magnification of jump discontinuities in $S_{abs}$ is 
anticipated since it is injective and provides the
necessary distinctness for different 
skyrmionic and consequently 
energy states. Therefore, it is being 
used in order to detect and 
explore possible energy barriers during the 
reversal process. Additionally, in \textbf{Fig.  \ref{fig:HzKxKu100skyrmionloop}} differentiation of shapes 
describing $S,S_{abs}$ reveals that there 
are local topological density $q_{Sk}$ 
regions-domains which sum up to zero and therefore 
mutually annihilate. 

Visualizations of micromagnetic configurations 
are shown in \textbf{Fig.	\ref{fig:HzKxSsnapshots}} 
for the representative selected external 
field values depicted in 
\textbf{Fig. \ref{fig:HzKxKu100skyrmionloop}} 
and designated as A, B, C, D. 
The actual topological density  $q_{Sk}$ defined in \textbf{Eq.  \ref{eq:chargedensity}} represents the "local" skyrmion 
number of the surface element and is also visualized 
in \textbf{Fig. \ref{fig:HzKxSsnapshots}}. 
This enriched representation gives
both qualitative and quantitative description of the actual
magnetization configuration. 
The existence of domains with augmented
local topological density  is observed
close to upper and lower corners of the nanoelement along
$y$-direction for A, B and D points. 
In addition, an elliptical domain can be
observed along the $x$-axis which is also the
direction of the in plane 
MCA $Ku\mathrm{=100 kJ/m^3}$, at the center of 
the triangular base present in all characteristic 
points A, B, C, D. These three magnetic entities can 
be considered as incomplete skyrmions and are present for a 
wide range of external field values. 
In \textbf{Fig. \ref{fig:HzKxSsnapshots}} (point A) 
the aforementioned magnetic entities expose 
different magnetization circulations. Those in the corners have 
a counter clockwise circulation while 
the domain with augmented $q_{Sk}$ at the center 
of the triangle has a clockwise circulation. 
\begin{figure}[!t] 
\centering
		\includegraphics[totalheight=1.6in,angle=0]{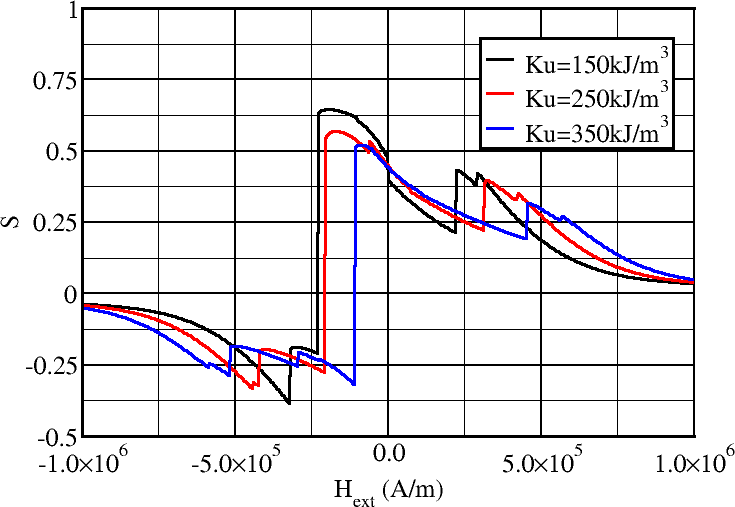}
		\caption{$S$ as a function of $H_{ext}$ for $K_u=\mathrm{150-350kJ/m^3}$ with MCA parallel to x-direction and $\mathbf{H}_{ext}//[001]$.} 
		\label{fig:HzKxskyrmionVSHanK}
\end{figure}

\begin{figure}[!t] 
\centering
		\includegraphics[totalheight=1.6in,angle=0]{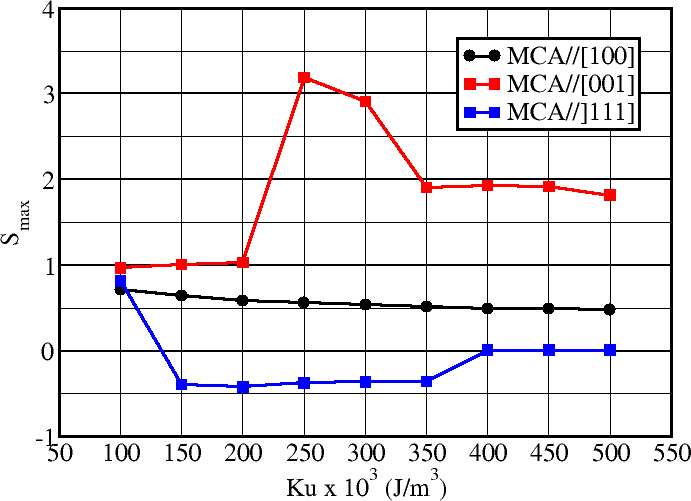}
		\caption{$S_{max}$ as a function  $K_u$ for MCA being in plane (//x), perpendicular (//z) and parallel to [111] direction with respect to the surface base of the nanoelement with the external field  $\mathbf{H}_{ext}//[001]$.}
		\label{fig:HzKskyrmionmax}
\end{figure}

In \textbf{Fig. \ref{fig:HzKxskyrmionVSHanK}}, skyrmion number 
$S$ as a function of
the external field $H_{ext}$ for different values of $Ku$ is shown.
As the magnitude of MCA increases gradually starting 
from $Ku\mathrm{=100kJ/m^3}$ a
similar behavior with \textbf{Fig. \ref{fig:HzKxKu100skyrmionloop}}
is observed regarding the qualitative and quantitative characteristics
of $S$.  In the cases of $Ku\mathrm{=150,250,350kJ/m^3}$ 
the skyrmion number 
attains lower values as $Ku$ increases exhibiting jump discontinuities. 
The calculated maximum (or negative minimum) values denoted as $S_{max}$ for the aforementioned
cases do not exceed the
value $S_{max}=0.7$ for $Ku\mathrm{=100kJ/m^3}$ represented in \textbf{Fig. \ref{fig:HzKxskyrmionVSHanK}}. 
For $Ku$=200 and $\mathrm{300kJ/m^3}$
$S$ attains maximum values 0.55--0.50 respectively but at different external magnetic field 
values $H_{ext}$. It is evident that the
magnetization reversal mechanism necessitates the formation of
incomplete skyrmions ($0.5<S<1$), vortex-like states ($S=0.5$)
not only for low $Ku$=100 and $\mathrm{150kJ/m^3}$  values but
also for the considered as intermediate values of $Ku\mathrm{=250,350 kJ/m^3}$
(\textbf{Fig. \ref{fig:HzKskyrmionmax}}).
Further increase of MCA's values to $Ku\mathrm{=400,450,500kJ/m^3}$
 expose similar reversal characteristics 
 with $Smax$ establishing 
a plateau region at 0.5 shown in \textbf{Fig. \ref{fig:HzKskyrmionmax}} 
characteristic for vortex like  reversal process.   

The discontinuities detected in case of $Ku\mathrm{=100 kJ/m^3}$
are present not only for different values of $Ku$  
but for different orientation of the MCA with respect to the
surface of the nanoelement e.g. parallel to z-direction. 
They need further clarification 
and  can be associated to the rich
energetic environment having 
contributions from demagnetization $E_{demag}$,
exchange $E_{exchange}$ and anisotropic $E_{anis}$ 
energies which can be computed for the systems
studied in the present work.
Complicated and rich energetic landscapes on the surface of the nanoelement are anticipated. The skyrmion formation is 
related to the interplay of these energetic
contributions. 

\begin{figure}[!t]
\centering
	\includegraphics[totalheight=1.6in,angle=0]{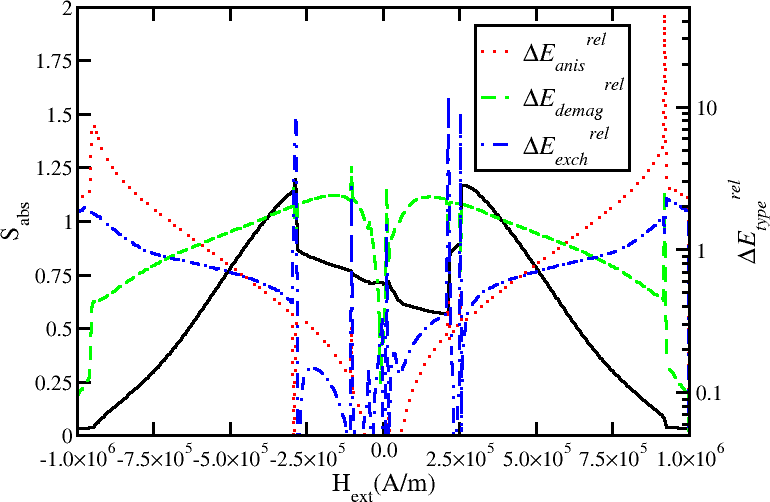}
	\caption{Relative values in $\%$ percentages for energetical types of demagnitization, exchange and anisotropic for $K_u=100$ for MCA parallel to [100].} 
	\label{fig:HzKxrelenrg}
\end{figure}

In order to measure the effect of each 
individual energy type on the micromagnetic 
configuration during the
magnetization reversal process the 
absolute relative energy difference 
$\Delta E_{type}^{rel}=|\frac{E_{type}^{i+1}-E_{type}^{i}}{E_{type}^{i}}|\times
100(\%)$ (where $type$ stands for $anis,exch,demag$) 
is computed between the consecutive external magnetic
field values $H_{ext}^{i},H_{ext}^{i+1}$. 
As mentioned in \textbf{Section \ref{sec:model}} 
the external magnetic field
step between consecutive field values is $\delta H_{ext}\mathrm{=4kA/m}$. 
The values of the relative differences 
of anisotropy, demagnitization 
and exchange energies are shown in
\textbf{Fig. \ref{fig:HzKxrelenrg} } as functions of $H_{ext}$ for MCA values of $Ku\mathrm{=100,150kJ/m^3}$ 
superimposed with $S_{abs}$.
It is clear that even in the first 
steps of the magnetization reversal
around values $\mathrm{8.5-9.5 \times 10^5A/m}$ 
of the external field for
$Ku\mathrm{=100kJ/m^3}$ small steps of the 
skyrmion number are induced by jump
discontinuities on the relative energies 
and vice versa.  These jump 
discontinuities  represent the actual energy barriers 
are pronounced for all types of energies at the
beginning of the reversal process with $E_{anis}$ 
being the more prominent compared to $E_{demag},E_{exch}$. The magnetic system and its associated energy should overcome a significant energy barrier in order to start the reversal process in the confined triangular prismatic nanomagnet. In particular, $\Delta
E_{anis}^{rel}$ attains values close to 42\%. Further reduction of the external magnetic field shows continuous behavior up to the value of $H_{ext}\mathrm{=2.5\times 10^5A/m}$ for $S_{abs}$.  The decrease of the external field $H_{ext}$ drives the continuous and gradual formation
of an incomplete skyrmion ($S < 1$).  This continuous behaviour is dictated from the continuous behaviour of the total magnetization energy
and its energetic components $E_{anis},E_{demag},E_{exch}$.  A new energy barrier is observed for $S_{abs}$ around $H_{ext} \mathrm{< 2.5 \times 10^5A/m}$ for the exchange energy $E_{exch}$ not followed by the
other calculated energies. The value $\Delta E_{exch}^{rel}$ of this discontinuity is close to 10\%.  It is clear that this sharp change at $E_{exch}$ induces the incomplete skyrmion discontinuity not only for
$H_{ext} \mathrm{< 2.5 \times 10^5A/m}$ but also for the symmetric with
respect $H_{ext}\mathrm{ =0A/m}$ second maximum 
around $H_{ext} \mathrm{=- 2.5 \times 10^5A/m}$. 

In particular, for external fields 
in the vicinity 
of $H_{ext} \mathrm{=0\; A/m}$ 
a steep descent is evident 
for $\Delta E_{demag}$.  
Energy component $\Delta E_{anis}$ shows a smoother behavior between the first and last 
discontinuities having a declining behavior in the first half of the magnetization reversal process where the 
external field approaches zero value. A continuous increase of 
the relative change $\Delta E_{anis}$ is
profound in the second half and final stage 
of the reversal process. Relative energy difference 
$\Delta E_{exch}$ exhibits 
an interesting non-continuous behavior having a significant 
number of sharp discontinuities in the 
range of $H_{ext} \mathrm{=(-2.5\; to\; +2.5) \times
10^5A/m}$ external field values.  
Calculation of energies and relative energies differences 
for varied $Ku$  showed similar qualitative and quantitative 
characteristics with the case of $Ku\mathrm{=100kJ/m^3}$. 
Vortex like or incomplete skyrmion states are triggered 
by abrupt energy changes and energy barrier crossings. 

Further calculations have performed by changing the direction of MCA from $x$ to $y$ direction. This directional change of
MCA gives similar physical results regarding the formation of
vortices despite the fact that MCA's direction remains in plane. This situation with MCA along y-direction  differs from
MCA lying along x-direction only with respect to the subtle effects of the edge curling on the nucleation process.
\subsection{Perpendicular MCA}
\begin{figure}[!t]
	\centering
\includegraphics[totalheight=1.6in,angle=0]{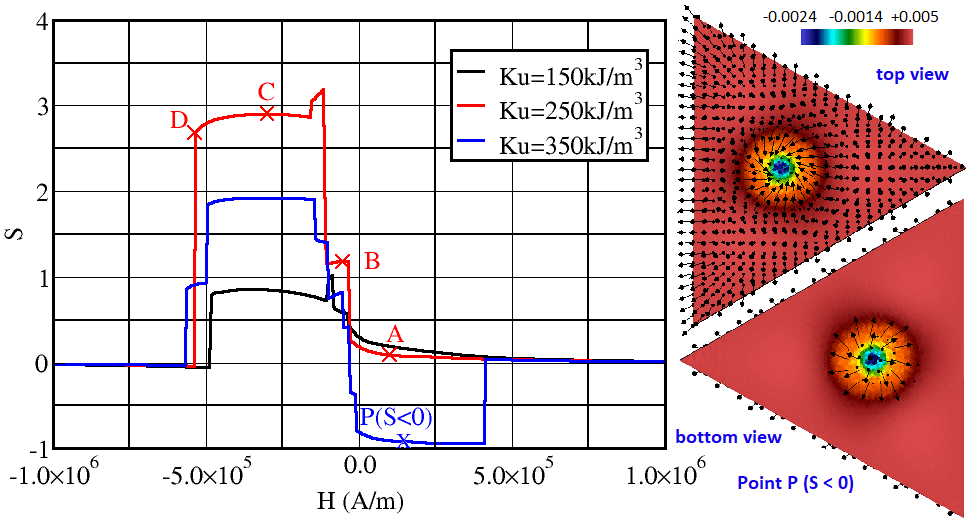}
\caption{ $S$ as a function of $H_{ext}$ for $Ku\mathrm{=150,250,350kJ/m^3}$ with MCA running parallel to z-direction alongside with the micromagnetic configuration in top and bottom view (rotated with respect to top view by 60 degrees)  for $Ku\mathrm{=350kJ/m^3}$ 
when skyrmion number attains negative values ($S=-1$) at representative point P.}
\label{fig:HzKzskyrmionVSHanK}
\end{figure}

It is challenging to investigate the effects associated 
with MCA when is set to $z$-direction 
which is the direction normal 
to the basis surface of the triangular FePt nanoelement. 
As the vortex-like formations in this system arise from the 
competition of exchange and magnetostatic energies it 
is expected that these phenomena will be more pronounced 
when there is a perpendicular MCA which leads 
to strong demagnetizing fields.
The numerical calculations showed that the 
dependence of $S$ for the lowest
MCA value $Ku\mathrm{=100kJ/m^3}$ is quite 
similar to the case of in-plane MCA
directions exhibiting the same magnetization reversal process.  
The magnetic system case with $Ku\mathrm{=150kJ/m^3}$  is 
presented in \textbf{Fig. \ref{fig:HzKzskyrmionVSHanK}}. 
Since the $Ku$ value is not high enough to give perpendicular 
anisotropy as the $H_{ext}$ is reduced the system departs from 
saturation through a series of 
topologically non-trivial formations 
giving a gradual increase of the skyrmion 
number which attains the maximum
value $S \simeq 0.9$ in a plateau region between $H_{ext}\mathrm{=-2.5\times 10^5A/m}$ and $H_{ext}\mathrm{=-5 \times 10^5A/m}$ is evident.  The aforementioned
gradual increase of the skyrmion number $S$  
for MCA parallel to z-direction for 
all $Ku$ values studied starts at lower 
external field values compared to systems having in
plane MCA direction. This external field values 
retardation can be attributed 
to the higher energy barrier needed to 
overcome in order to start  
the reversal process through skyrmion formation. 
The critical field signaling 
skyrmion formation for the different 
$Ku\mathrm{=150,250,350kJ/m^3}$ values for MCA parallel to z-direction is located around $H_{ext}\mathrm{=3.5 \times
10^5A/m}$ (\textbf{Fig. \ref{fig:HzKzskyrmionVSHanK}}) and is considerably lower compared to the external field 
value of $H_{ext}\mathrm{=7.5 \times 10^5A/m}$ for in-plane MCA (\textbf{Fig. \ref{fig:HzKxKu100skyrmionloop}}).
\begin{figure}[!t]
\centering
\includegraphics[totalheight=3.0in,angle=0]{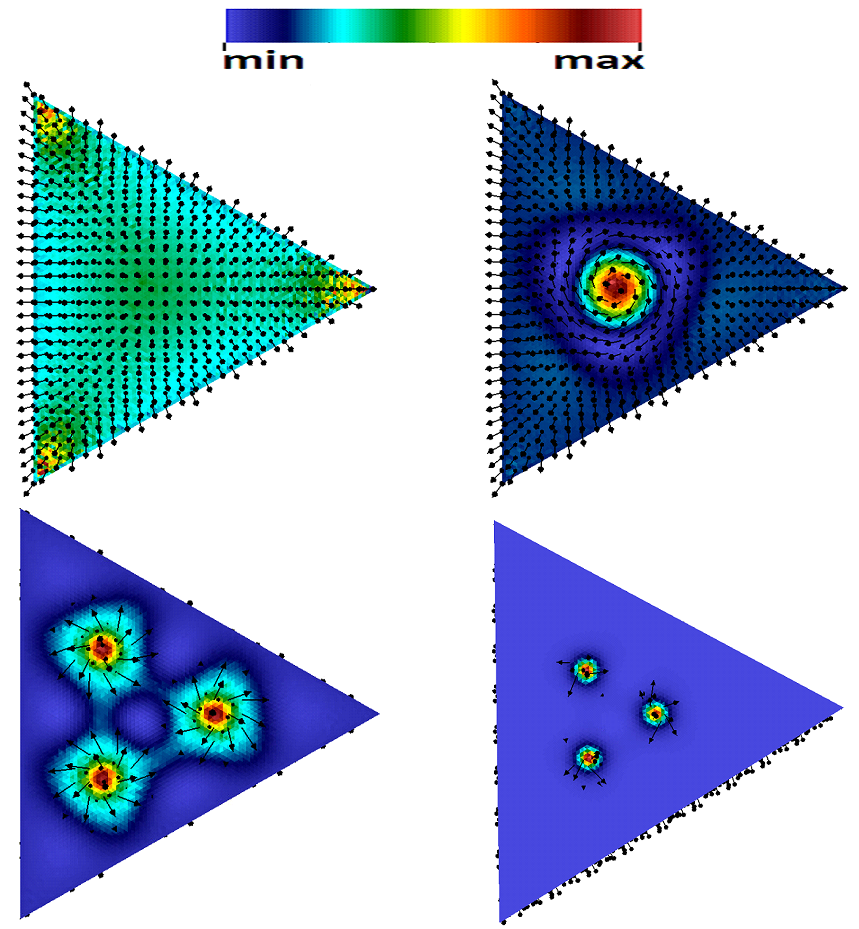}
\caption{Top view of micromagnetic configuration showing topological density $q_{Sk}$ superimposed with the local magnetization for $Ku\mathrm{=250kJ/m^3}$. Different diagrams depict the four points A, B, C, D of Fig. \ref{fig:HzKzskyrmionVSHanK} (from left to right and from top to bottom) describing the reversal process. The actual value ranges of $q_{Sk}/nm^{-2}$ are represented in color bars with 
maximum and minimum values respectively:
left top (A  $\max:3.419\times 10^{-5}\; -\; \min:-1.117\times 10^{-6}$) , right top 
	(B  $\max:0.00063\; -\; \min:-0.00012$ ), left down (C  $\max:0.00346\; -\; \min:-0.00020$), right down	(D  $\max:0.00063\; -\; \min:-0.00012$).}
\label{fig:HzKzsnapshotsVSHanK}
\end{figure}

The behavior is strikingly different 
when MCA attains the values 
of $Ku\mathrm{=250, 350kJ/m^3}$ depicted also 
in  \textbf{Fig. \ref{fig:HzKzskyrmionVSHanK}}. For the case where $Ku\mathrm{=350kJ/m^3}$ skyrmionic configurations having 
negative values emerge for external field values lower than 
$5.0\times 10^5$A/m. The micromagnetic  configuration in \textbf{Fig. \ref{fig:HzKzskyrmionVSHanK}} hosts one skyrmion
located at the center of the nanoparticle. A counterclockwise 
circulation (positive circulation) can be identified on the top view of the nanoparticle. The central spins of the skyrmion point inwards (negative polarity) as clearly presented at the bottom view of the spin configuration. Consequently, chirality which  
is the product of circulation and polarity 
is negative dictating in this manner the negativity of the skyrmion number. 

For the particular case of $Ku\mathrm{=250kJ/m^3}$ regions hosting skyrmions are evident having 
skyrmion number values close to $S=3$ in a range of
applied external fields from 
$H_{ext}\mathrm{=-0.3\times 10^5A/m}$ to $\mathrm{- 5.3\times 10^5A/m}$. 
In \textbf{Fig. \ref{fig:HzKzsnapshotsVSHanK}} micromagnetic configurations with topological density $q_{Sk}$ superimposed with the local  magnetization vector are presented. 
Different diagrams depict four representative 
points A, B, C, D in four characteristic regions 
during the reversal process of the $S$ variation 
with the external field $H_{ext}$. 
In \textbf{Fig. \ref{fig:HzKzsnapshotsVSHanK}} 
(point A in \textbf{Fig. \ref{fig:HzKzskyrmionVSHanK}} for $Ku\mathrm{=250kJ/m^3}$) skyrmion is under development. 
Regions of augmented $q_{Sk}$ are evident exactly at the triangle's corners. Lower values of $q_{Sk}$ can be seen at the center of the triangular base of the nanoelement. 
Point B is located at a skyrmion number $S$ 
discontinuity with actual value close to $S=1$. The 
respective isosurfaces of $q_{Sk}$ for point B  
reveal the existence  of one skyrmion located at 
the center of the triangular prismatic nanoelement. 

Points C, D represent micromagnetic configurations hosting three skyrmions. The centers of the skyrmions define an equilateral triangle in both micromagnetic configurations. 
The essential difference of the skyrmions 
present in C and D points   
is their actual size.  Skyrmions of point C are larger compared 
to the respective three skyrmions in D although 
they have the same circular shapes and location. 
This can be attributed to the fact that while the sizes 
in skyrmions of C (\textbf{Fig. \ref{fig:HzKzsnapshotsVSHanK}} (C) ) are larger than those of D (\textbf{Fig. \ref{fig:HzKzsnapshotsVSHanK}} (D) ) the topological densities $q_{Sk}$ are higher in D
in order to compensate the lower skyrmion surface contributing to the total
triangular surface base integration giving in both cases C and D 
a total skyrmion number equal to 3. 
Representative values of the maximum $q_{Sk}$ on the surface element are $q_{Sk,C}=0.003462 < q_{Sk,D}=0.02181$.   

\begin{figure}[!t]
\centering
\includegraphics[totalheight=2.6in,angle=0]{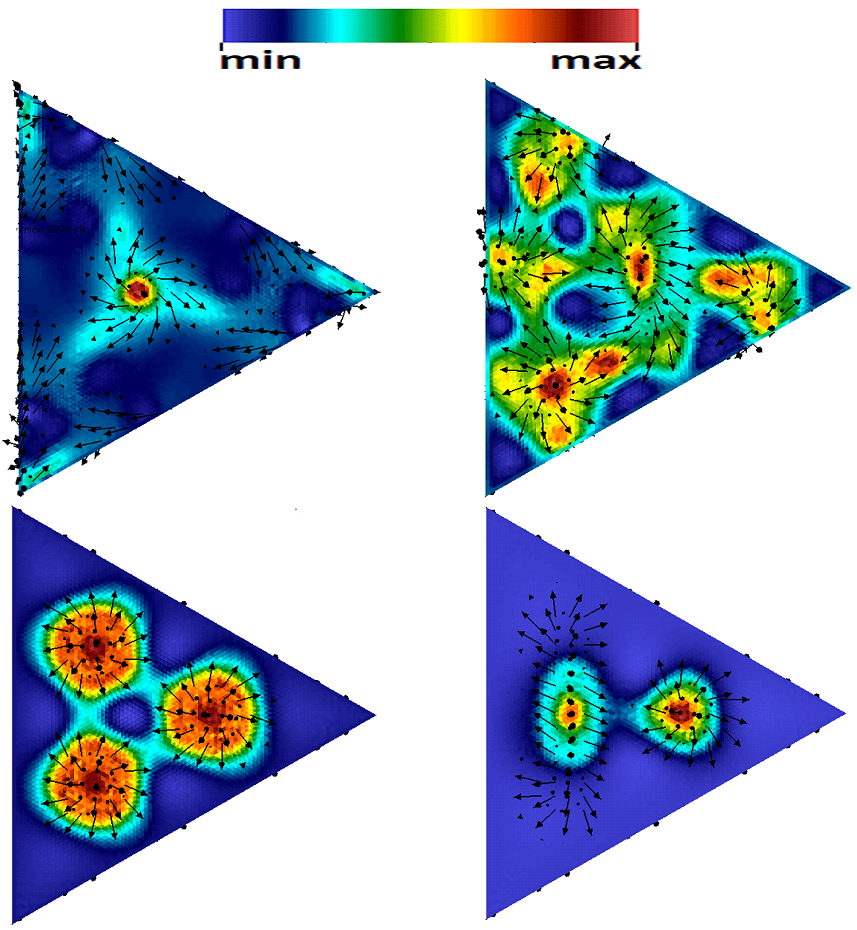}
	\caption{Top view of micromagnetic configurations referring to $S_{max}$  for $Ku\mathrm{=200,250,300,350kJ/m^3}$ (from left to right and from top to bottom) for MCA being parallel to z-direction. Magnetization shown with arrows superimposed with $q_{Sk}$. The actual value ranges of $q_{Sk}/\mathrm{nm^{-2}}$ are represented in color bars with maximum and minimum values respectively: 
left top ($Ku\mathrm{=200kJ/m^3}$, $max:0.00342\; -\; min:-0.00643$) , right top 
	($Ku\mathrm{=250kJ/m^3}$, $max:0.00175\; -\; min:-0.00043$ ), left down ($Ku\mathrm{=300kJ/m^3}$, $max:0.00135\; -\; min:-0.00020$), right down ($Ku\mathrm{=350kJ/m^3}$, $max:0.00322\; -\; min:-9.61\times 10^{-5}$).}
	\label{fig:HzKzSmaxsnapshots}
\end{figure}

In \textbf{Fig. \ref{fig:HzKskyrmionmax}} the maximum 
values of $S$  denoted as $S_{max}$ are reported 
for MCA parallel to z-direction. Moreover, 
representative configurations depicted for external field values 
corresponding to  $S_{max}$ showing $q_{Sk}$ superimposed with magnetization are presented in \textbf{Fig. \ref{fig:HzKzSmaxsnapshots}}. For the case where $Ku\mathrm{=200kJ/m^3}$ at the center of the triangular base a clear 
and complete skyrmion emerges. 
It has an almost perfect circular shape
with the higher values of $q_{Sk}$ located at 
the center mitigating when  
moving away but along the circle's radius. The magnetization vectors have a counter-clockwise circulation around the skyrmion. 
In case where 
$Ku\mathrm{=250kJ/m^3}$ and $S_{max} \approx 3.25$ extended regions located at 
the triangle's vertices running along the edges of the triangular 
base are evident. This magnetic configuration significantly differs 
from the configuration having three well developed  skyrmions defining segregated and distinct regions 
presented in \textbf{Fig. \ref{fig:HzKzsnapshotsVSHanK}} (point C). 
These multiple skyrmions or clusters of skyrmions  
on the surface of
the triangular disk are evident and the actual calculation of
$S$ gives values during the reversal process in 
the range of $1 \le S < 4$ for all MCA values studied having perpendicular
anisotropy. The maximum value $S_{max}=3$  for 
the case of $Ku\mathrm{=300kJ/m^3}$ is also depicted in \textbf{Fig.  \ref{fig:HzKzSmaxsnapshots}} exposing three skyrmions located 
in extended circular regions.

The increase of the MCA value from $Ku\mathrm{=150kJ/m^3}$ to $\mathrm{500kJ/m^3}$ 
plays a dramatic effect in
the magnetization reversal process and the creation of skyrmion
regions. Particularly, in the cases where $Ku$ attains values beyond  
$\mathrm{300kJ/m^3}$  reaching  $\mathrm{500kJ/m^3}$ the maximum of 
skyrmion number $S$ establishes a plateau 
at $S=2$ indicative of the formation of two skyrmions on the 
surface of the nanoelement. The case of  $Ku\mathrm{=350kJ/m^3}$ where  
$S_{max}=2$ is representative and is shown in \textbf{Fig.  \ref{fig:HzKzSmaxsnapshots}}. The two skyrmions are exactly located along the height of the triangle (parallel to x-direction) having different shapes resembling different perturbations of the circular shape. 
The formation and actual detection of skyrmions 
can be revealed  by  visualizing the 
actual reversal process. A quantitative 
and coherent picture can be provided  
by the calculation of $S$.

\begin{figure}[!b]
\centering
	\includegraphics[totalheight=1.6in,angle=0]{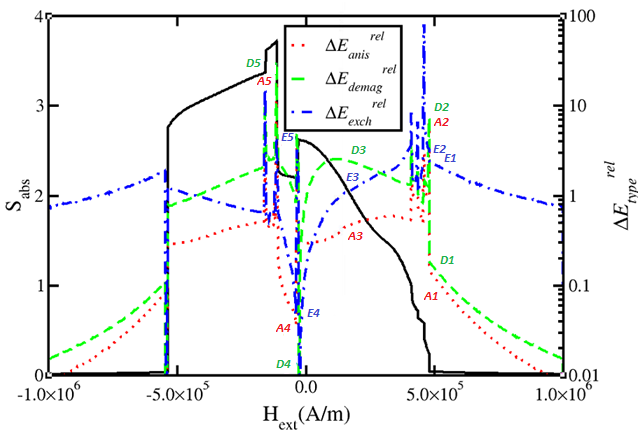}
	\caption{Values in $\%$ percentages for relative energy differences of demagnitization, exchange and anisotropic for $Ku\mathrm{=250kJ/m^3}$ for MCA parallel to [001].} 
	\label{fig:HzKzrelenrg}
\end{figure}

Computations of relative energy differences 
accompanied by $S_{abs}$ are depicted in 
\textbf{Fig. \ref{fig:HzKzrelenrg}} 
for $Ku\mathrm{=250kJ/m^3}$ which is the 
case showed the formation of three 
skyrmions. 
As it is already commented the first steps of 
the magnetization reversal for MCA normal 
to nanoelement's surface start 
at external field values $H_{ext}$ significantly lower 
($<5 \times 10^5$A/m) compared to the case of in-plane MCA. Following $S_{abs}$ 
and relative energies 
variation with respect to external field, 
energy barriers  are present 
around $H_{ext}\mathrm{=5 \times 10^5A/m}$ for 
all the  components of energy.  
Particularly in \textbf{Fig. \ref{fig:HzKzrelenrg}} 
for $Ku\mathrm{=250kJ/m^3}$, $\Delta E_{exch}^{rel}$ is close $90\%$ and 
is the most prominent jump discontinuity compared to 
$\Delta E_{anis}^{rel}$, $\Delta E_{demag}^{rel}$ whose jumps are around
$10\%$. It should also noted that $E_{anis}^{rel}$ exhibits a  $90\%$ jump, which is twice as much
as the one observed in the case of in plane anisotropy at $Ku\mathrm{=100kJ/m^3}$ which was calculated to 42\%. 
The reduction of the external magnetic field affects in a  
continuous manner the different magnetic energies  up to the value of $H_{ext}=0$ A/m.  In the vicinity of the zero for negative external fields 
new jump discontinuities are evident for all energies. The decrease of external field $H_{ext}$ drives the continuous and gradual formation followed by energetic and therefore skyrmionic discontinuity events.
Complete and incomplete skyrmions are present having values 
$1< S < 4$.   Rich energy patterns and textures are clear having similar behaviour not only for  $Ku\mathrm{=250kJ/m^3}$  but for all $Ku$ values studied when MCA is normal to the nanoelements' base. 

\begin{figure*}[!t]
	\centering
	\includegraphics[totalheight=6.6in,angle=0]{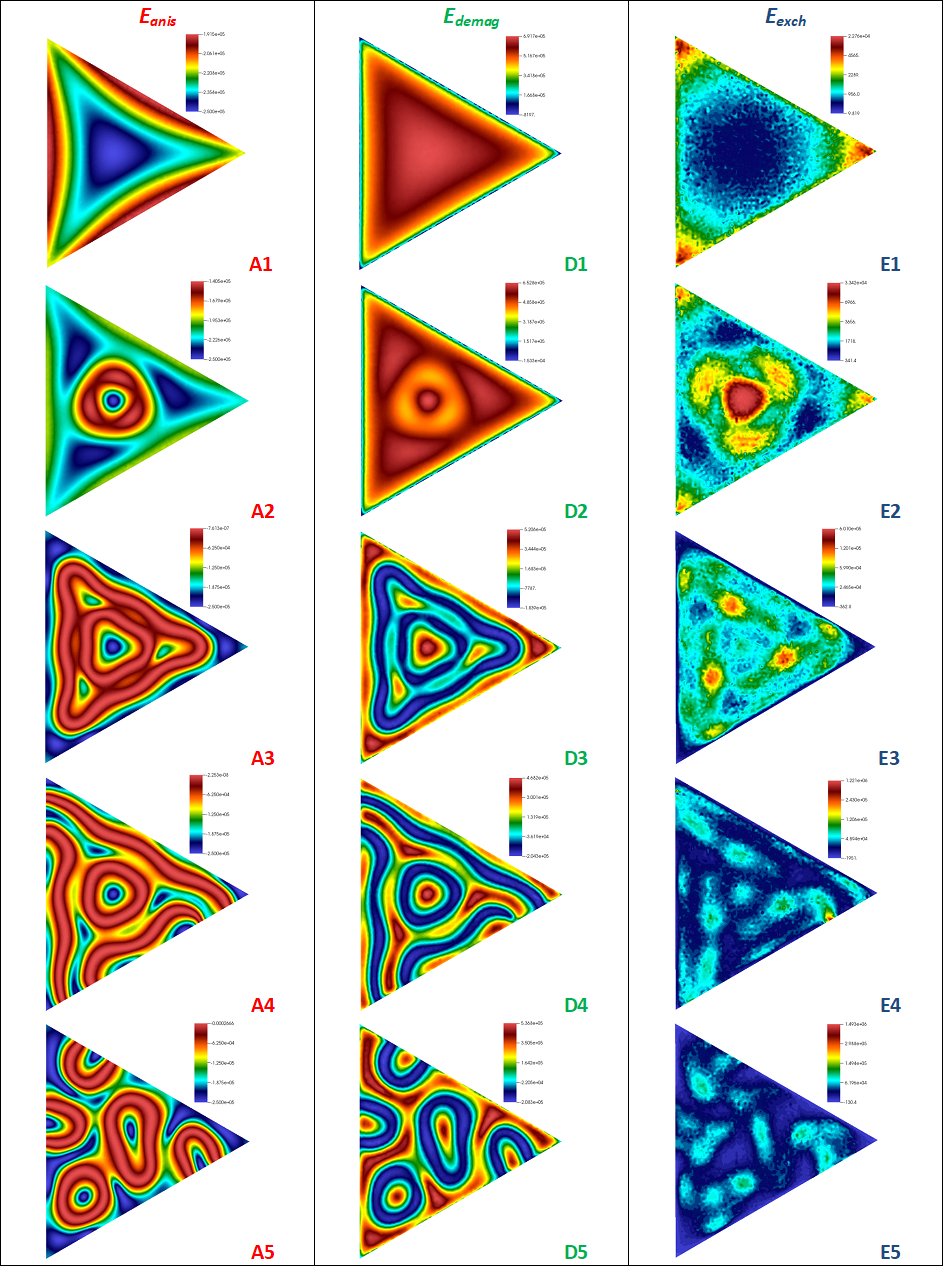}
	\caption{Demagnitization, exchange, and anisotropic energies for $Ku\mathrm{=250kJ/m^3}$ for selected points during the reversal process chosen from \textbf{Fig. \ref{fig:HzKzrelenrg}}. } 
	\label{fig:HzKzenergyphotos}
\end{figure*}
\textbf{Fig. \ref{fig:HzKzenergyphotos}} represents 
the relative energies for characteristic points selected from  \textbf{Fig. \ref{fig:HzKzrelenrg}} at $Ku\mathrm{=250kJ/m^3}$. Energies  $E_{anis},E_{exch},E_{demag}$ are 
grouped and shown in  
five different points.  The grouped  
energies $\{A_i,D_i,E_i\}$ associated 
with point $i$ ($i=1,..,5$) 
are depicted at the same or approximately 
the same $H_{ext,i}$ value. 
At point $A1$ before crossing  the energy barrier, $E_{anis}$ develops an inner triangular region with  values lower compared to the maximum values located at the 
sides of the nanoelement's base. At $A_2$ point where the barrier is being crossed and skyrmion has been formed 
the energy distribution follows 
the skyrmion's  location with increasing  values of $E_{anis}$ 
radially moving from the center towards the 
outer circular domain of skyrmion. Energy isosurfaces 
for point $A3$ expose three new developed regions in the vicinity of the nanoelement's base vertices. Point $A4$ represents a local  energy mimimum of $E_{anis}$ in which around the 
skyrmion domain at the center 
 well developed stripe-like energy domains running parallel 
to nanoelement's edges are evident. Point $A5$ is 
the location where the highest energy barrier is detected. From the contour plot is evident that the energy follows exactly the location of the multiple skyrmions formed.  Calculation of $E_{demag}$ provides the same qualitative behavior with the subtle difference on point $D1$ at almost identical $H_{ext}$ value with $A1$ where $E_{demag}$ has
a uniform distribution on the nanoelement's base. 
The calculated exchange energy $E_{exch}$ 
distribution follows the formation process of skyrmions. 
At point $E1$ high 
energy regions are located at
the vertices of the nanoelement in contrast 
to $E_{anis}$. The competition between these 
energies and energy barrier crossings 
gives birth to skyrmion magnetic configurations. 
%
%
\subsection{MCA parallel to [111]}
In addition to MCA lying on the surface or being normal 
to the surface of the nanoelement the magneto-crystalline anisotropy 
is set parallel to [111] direction.The latter case is 
of interest as in many instances FePt and CoPt films 
tend to grow with their $[111]$ crystallographic directions 
resulting an angle of $54.7^{o}$ to the film normal \cite{Alexandrakis:2009}.
\begin{figure}[!t]
	\centering
		\includegraphics[totalheight=1.6in,angle=0]{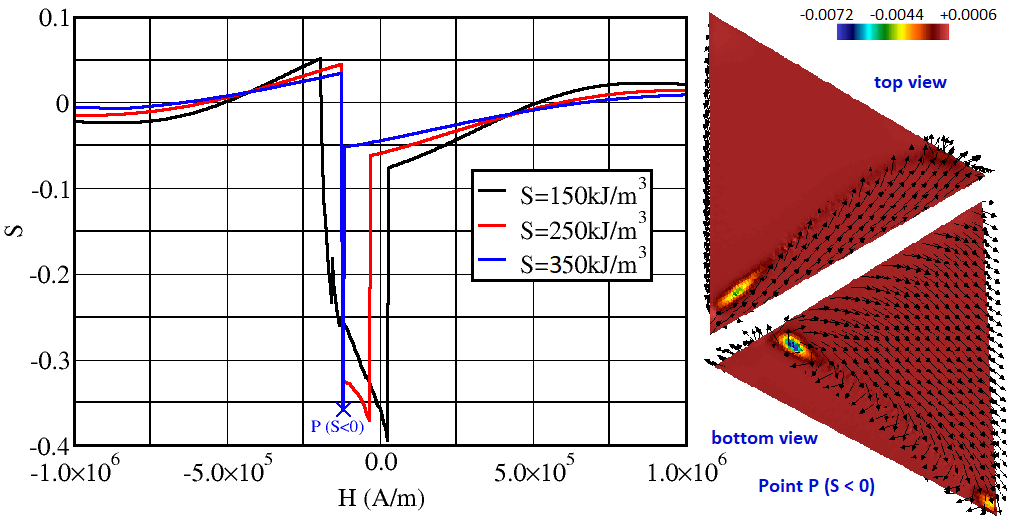}
		\caption{ $S$ as a function of $H_{ext}$ for $Ku\mathrm{=150,250,350kJ/m^3}$ with MCA running parallel to [111] direction alongside with the micromagnetic configuration in top and bottom view (rotated with respect to top view by 60 degrees) for  representative point P with $S<0$ at $Ku\mathrm{=350kJ/m^3}$. } 
		\label{fig:HzK111_funKu}
\end{figure}
The numerical simulation results are presented for $S$
in \textbf{Fig. \ref{fig:HzKxtopbottom}} as a function of $Ku$ 
fixing external field value at $H_{ext}$=0A/m. 
At $Ku\mathrm{=100kJ/m^3}$ skyrmion
number is close to value $S=0.7$. By 
increasing MCA up to $Ku\mathrm{=250kJ/m^3}$ 
skyrmion number reduces attaining negative but non-zero 
values for external field $H_{ext}$=0A/m.  
This is an interesting fact and significantly differentiates 
compared to the cases where MCA is parallel 
to x and z (in the majority of $Ku$ 
values studied) directions where positive 
skyrmion numbers are obtained. The negative values originate 
from the negative chirality (circulation $\times$ polarity)
of the high skyrmion density $q_{Sk}$ region  that can be seen in \textbf{Fig. \ref{fig:HzK111_funKu}}. 

The maximum skyrmion number value $S_{max}$ as 
a function of external field 
for different MCA values 
is represented in \textbf{Fig. \ref{fig:HzKskyrmionmax}}.
Topological quantity $S_{max}$ attains negative 
values in contrast to the cases where MCA is in-plane or 
perpendicular to the nanoelement's base. For $S_{max}<0$ the 
respective MCA value range is $Ku\mathrm{=150-350kJ/m^3}$.
Calculation of the skyrmion number at different external field values 
for the representative $Ku\mathrm{=150,250,350kJ/m^3}$ 
values are shown in
\textbf{Fig. \ref{fig:HzK111_funKu}}. Evident 
is the manifestation of micromagnetic configurations 
mainly having negative skyrmion numbers
for external fields ranging from $H_{ext}\mathrm{=5\times 10^5A/m}$ 
to approximately $H_{ext}\mathrm{=-2 \times 10^5A/m}$.
The minimum value of $S$ is observed for the 
case of $Ku\mathrm{=150kJ/m^3}$ and 
is close to $S=-0.4$. For the three different $Ku$ values 
shown in \textbf{Fig. \ref{fig:HzK111_funKu}} 
the crossover to negative $S$ happens at $H_{ext}\mathrm{=5\times 10^5A/m}$. Also the maximum antivortex states are 
located at the vicinity of zero external 
magnetic field which significantly 
differentiate compared to the previous studied cases. 

\section*{Conclusions}
The topological invariant known as skyrmion number $S$
has been calculated for FePt triangular prism nanonelements with
different directions and magnitude of MCA for varying magnitude 
of the external magnetic field $H_{ext}$. The direction 
of the external field $\mathbf{H}_{ext}$ was normal to nanoelement's 
surface in all conducted numerical simulations. 
Magnetization configurations during reversal process were 
studied. It was possible to explore the
formation of skyrmionic regions and qualitatively-quantitatively
characterize them by a variety of calculated properties such as the
skyrmion number, the different contributions 
on the magnetic energy such as
demagnetization energy,  exchange energy and uniaxial anisotropy
energy and visualization of the micromagnetic configurations. 
Skyrmionic magnetic configurations have been detected
in high symmetry positions with respect to 
the geometry of the triangular prism having skyrmion number greater than one ( $S> 1$ )
 for the case where MCA was normal 
to nanoelement's base. For  MCA attaining values 
between $Ku\mathrm{=200-500kJ/m^3}$ three distinct 
skyrmions are formed and
persist for a range of external fields.

In conjunction with previous studies it is clear that magnetic
skyrmions can be produced in a wide range of external fields 
just by tuning  MCA's magnitude and direction, even in the absence of chiral interactions such as
Dzyaloshinsky-Moriya. It is challenging to 
extent the numerical simulations including thermal effects 
in the form of Brownian term in LLG
equation in order to investigate the formation of skyrmions at room temperature.

\section*{Acknowledgments}
V.D. Stavrou would like to thank the State Scholarship Foundation of Greece (IKY) for the financial support under the scholarship grant (appl. no.14386). We would like to thank
Mr. Costas Dimakopoulos for his technical support. Computations have been performed at the Laboratory of Mathematical
Modeling and Scientific Computing of the Materials Science Department
of the University of Ioannina.

\bibliographystyle{plain}
\bibliography{skyrmion}

\begin{thebibliography}{10}

\bibitem{Alexandrakis:2009}
V.~Alexandrakis, D.~Niarchos, M.~Wolff, and I.Panagiotopoulos.
\newblock Magnetization reversal in \text{CoPt} (111) hard/soft bilayers.
\newblock {\em Journal of Applied Physics}, 105:063908, 2009.

\bibitem{Hovorka:2017}
Marijan Beg, Maximilian Albert, Marc-Antonio Bisotti, David
  {Cort\'es-Ortu\~no}, Weiwei Wang, Rebecca Carey, Mark Vousden, Ondrej
  Hovorka, Chiara Ciccarelli, Charles~S. Spencer, Christopher~H. Marrows, and
  Hans Fangohr.
\newblock Dynamics of skyrmionic states in confined helimagnetic
  nanostructures.
\newblock {\em Phys. Rev. B}, 95:014433, Jan 2017.

\bibitem{Bogdanov:1994}
A.~Bogdanov.
\newblock Thermodynamically stable magnetic vortex states in magnetic crystals.
\newblock {\em Journal of Magnetism and Magnetic Materials}, 138(3):255--269,
  1994.

\bibitem{Bogdanov:1995}
A.~Bogdanov and A.~Hubert.
\newblock Thermodynamically stable magnetic vortex states in magnetic crystals.
\newblock {\em JETP Lett.}, 62:247--251, 1995.

\bibitem{Buttner:2018}
Felix B\"uttner, Ivan Lemesh, and Geoffrey S.~D. Beach.
\newblock Theory of isolated magnetic skyrmions: From fundamentals to room
  temperature applications.
\newblock {\em Scientific Reports}, 8:4464, 2018.

\bibitem{Dai:2013}
Y.~Y. Dai, H.~Wang, P.~Tao, T.~Yang, W.~J. Ren, and Z.~D. Zhang.
\newblock Skyrmion ground state and gyration of skyrmions in magnetic nanodisks
  without the dzyaloshinsky-moriya interaction.
\newblock {\em Phys. Rev. B}, 88:054403, Aug 2013.

\bibitem{Fert:2017}
Albert Fert, Nicolas Reyren, and Vincent Cros.
\newblock {Magnetic skyrmions: advances in physics and potential applications}.
\newblock {\em Nature Reviews: Materials}, 2:17031, 2017.

\bibitem{Nmag:2007}
T.~Fischbacher, M.~Franchin, G.~Bordignon, and H.~Fangohr.
\newblock A systematic approach to multiphysics extensions of
  finite-element-based micromagnetic simulations: Nmag.
\newblock {\em IEEE Trans. Magn.}, 43:2896--2898, 2007.

\bibitem{FangohrNov:2015}
T.~Fischbacher, M.~Franchin, G.~Bordignon, and H.~Fangohr.
\newblock Ground state search, hysteretic behaviour, and reversal mechanism of
  skyrmionic textures in confined helimagnetic nanostructures.
\newblock {\em Scientific Reports}, 5:6784, 2015.

\bibitem{Guslienko:2015}
Konstantin~Y. Guslienko.
\newblock {Skyrmion State Stability in Magnetic Nanodots with Perpendicular
  Anisotropy}.
\newblock {\em IEEE MAGNETICS LETTERS}, 6:4000104, 2015.

\bibitem{Hanneken:2015}
Christian Hanneken, Fabian Otte, Andr{\'e} Kubetzka, Bertrand Dup{\'e}, Niklas
  Romming, Kirsten von Bergmann, Roland Wiesendanger, and Stefan Heinze.
\newblock Electrical detection of magnetic skyrmions by tunnelling
  non-collinear magnetoresistance.
\newblock {\em Nature Nanotechnology}, 10:1039--1042, 2015.

\bibitem{He:2017}
Zhezhi He, Shaahin Angizi, and Deliang Fan.
\newblock Current-induced dynamics of multiple skyrmions with domain-wall pair
  and skyrmion-based majority gate design.
\newblock {\em IEEE MAGNETICS LETTERS}, 8:4305705, 2017.

\bibitem{Heinze:2011}
Stefan Heinze, Kirsten von Bergmann, Matthias Menzel, Jens Brede, Andre
  Kubetzka, Roland Wiesendanger, Gustav Bihlmayer, and Stefan Bl{\tiny }ugel.
\newblock Spontaneous atomic-scale magnetic skyrmion lattice in two dimensions.
\newblock {\em Nature Physics}, 7:713--718, 2011.

\bibitem{Iwasaki:2014}
Junichi Iwasaki, Aron~J. Beekman, and Naoto Nagaosa.
\newblock Theory of magnon-skyrmion scattering in chiral magnets.
\newblock {\em Phys. Rev. B}, 89:064412, Feb 2014.

\bibitem{Jaafar:2010}
M.~Jaafar, R.~Yanes, D.~Perez~de Lara, O.~Chubykalo-Fesenko, A.~Asenjo, E.~M.
  Gonzalez, J.~V. Anguita, M.~Vazquez, and J.~L. Vicent.
\newblock Control of the chirality and polarity of magnetic vortices in
  triangular nanodots.
\newblock {\em Phys. Rev. B}, 81:054439, Feb 2010.

\bibitem{Jeong:2001}
S.~Jeong, M.E. McHenry, and D.E. Laughlin.
\newblock Growth and characterization of {L10 FePt and CoPt} $<001>$ textured
  polycrystalline thin films.
\newblock {\em IEEE Transactions on Magnetics}, 37:1309--1311, 2001.

\bibitem{Jiang:2015}
Wanjun Jiang, Pramey Upadhyaya, Wei Zhang, Guoqiang Yu, M.~Benjamin
  Jungfleisch, Frank~Y. Fradin, John~E. Pearson, Yaroslav Tserkovnyak, Kang~L.
  Wang, Olle Heinonen, Suzanne G.~E. te~Velthuis, and Axel Hoffmann.
\newblock Blowing magnetic skyrmion bubbles.
\newblock {\em Science}, 349(6245):283--286, 2015.

\bibitem{Jin:2017}
C.~Jin, C.~Song, J.~Wang, H.~Xia, and J.~Wang.
\newblock Topological trajectories of a magnetic skyrmion with in-plane
  microwave magnetic field.
\newblock {\em Journal of Applied Physics}, 122:223901, 2017.

\bibitem{Sitte:2017}
K.Everschor-Sitte, M.~Sitte, T.~Valet, A.~Abanov, and J.~Sinova.
\newblock Skyrmion production on demand by homogeneous dc currents.
\newblock {\em New Journal of Physics}, page 092001, 2017.

\bibitem{Novosad:2010}
Dong-Hyun Kim, Elena~A. Rozhkova, Ilya~V. Ulasov, Samuel~D. Bader, Tijana Rajh,
  Maciej~S. Lesniak, and Valentyn Novosad.
\newblock {Biofunctionalized magnetic-vortex microdiscs for targeted
  cancer-cell destruction}.
\newblock {\em Nature Materials}, 9:165--171, 2010.

\bibitem{Komineas:1996}
S.~Komineas and N.~Papanicolaou.
\newblock Topology and dynamics in ferromagnetic media.
\newblock {\em Physica D}, 99:81--107, 1996.

\bibitem{Nagaosa:2016}
Wataru Koshibae and Naoto Nagaosa.
\newblock {Theory of antiskyrmions in magnets}.
\newblock {\em Scientific Reports}, 7:10542, 2016.

\bibitem{Nagaosa:2017}
Wataru Koshibae and Naoto Nagaosa.
\newblock {Theory of skyrmions in bilayer systems}.
\newblock {\em Scientific Reports}, 7:42645, 2017.

\bibitem{Li:2014}
J.~Li, A.~Tan, K.W. Moon, A.~Doran, M.A. Marcus, A.T. Young, E.~Arenholz,
  S.~Ma, R.F. Yang, C.~Hwang, and Z.Q. Qiu.
\newblock Tailoring the topology of an artificial magnetic skyrmion.
\newblock {\em Nature Communications}, 5(4704):1--14, November 2014.

\bibitem{Markou:2013}
A.~Markou, K.~G. Beltsios, L.~N. Gergidis, I.~Panagiotopoulos, T~Bakas,
  K.~Ellinas, A.~Tserepi, L.~Stoleriu, R.~Tanasa, and A.~Stancu.
\newblock {Magnetization reversal in triangular L1 0-FePt nanoislands}.
\newblock {\em Journal of Magnetism and Magnetic Materials}, 344:224--229,
  2013.

\bibitem{Ding:2014}
B.~F. Miao, L.~Sun, Y.~W. Wu, X.~D. Tao, X.~Xiong, Y.~Wen, R.~X. Cao, P.~Wang,
  D.~Wu, Q.~F. Zhan, B.~You, J.~Du, R.W. Li, and H.~F. Ding.
\newblock Experimental realization of two-dimensional artificial skyrmion
  crystals at room temperature.
\newblock {\em Phys. Rev. B}, 90:174411, 2014.

\bibitem{Moutafis:2016}
C.~Moreau-Luchaire, C.~Moutafis, N.~Reyren, J.~Sampaio, C.~A.~F. Vaz, N.~Van
  Horne, K.~Bouzehouane, K.~Garcia, C.~Deranlot, P.~Warnicke, P.~Wohlh\"uter,
  J.M. George, M.~Weigand, J.~Raabe, V.~Cros, and A.~Fert.
\newblock Additive interfacial chiral interaction in multilayers for
  stabilization of small individual skyrmions at room temperature.
\newblock {\em Nature Nanotechnology}, 11:444--448, 2016.

\bibitem{Muller:2017}
Jan M\"uller.
\newblock Magnetic skyrmions on a two-lane racetrack.
\newblock {\em New Journal of Physics}, 19:025002, 2017.

\bibitem{Nakatani:2016}
Y.~Nakatani, M.~Hayashi, S.~Kanai, S.~Fukami, and H.~Ohno.
\newblock Electric field control of skyrmions in magnetic nanodisks.
\newblock {\em Applied Physics Letters}, 108:152403, 2016.

\bibitem{NETGEN:2007}
{NETGEN}.

\bibitem{Okamoto:2002}
S.~Okamoto, N.~Kikuchi, O.~Kitakami, T.~Miyazaki, Y.~Shimada, and K.~Fukamich.
\newblock Chemical-order-dependent magnetic anisotropy and exchange stiffness
  constant of \text{FePt} (001) epitaxial films.
\newblock {\em Phys. Rev. B}, 66:024413, 2002.

\bibitem{Fangohr:2018}
Ryan~A. Pepper, Marijan Beg, David {Cort\'es-Ortu\~no}, ￼~Thomas Kluyver,
  Marc-Antonio Bisotti, Rebecca Carey, ￼~Mark Vousden, Maximilian~Albert
  and￼ Weiwei~Wang, Ondrej Hovorka, and Hans Fangohr.
\newblock Skyrmion states in thin confined polygonal nanostructures.
\newblock {\em Journal of Applied Physics}, 123:093903, 2018.

\bibitem{Romming:2013}
N.~Romming, C.~Hanneken, M.~Menzel, J.~E. Bickel, B.~Wolter, K.~von Bergmann,
  A.~Kubetzka, and R.~Wiesendanger.
\newblock Writing and deleting single magnetic skyrmions.
\newblock {\em Science}, 341:636--639, 2013.

\bibitem{Romming:2015}
N.~Romming, A.~Kubetzka, C.~Hanneken, K.~von Bergmann, and R.~Wiesendanger.
\newblock Field-dependent size and shape of single magnetic skyrmions.
\newblock {\em Physical Review Letters}, 114:177203, 2015.

\bibitem{Sampaio:2013}
J.~Sampaio, V.~Cros, S.~Rohart, A.~Thiaville, and A.~Fert.
\newblock Nucleation, stability and current-induced motion of isolated magnetic
  skyrmions in nanostructures.
\newblock {\em Nature Nanotechnology}, 8:839--844, 2013.

\bibitem{Sapozhnikov:2015}
M.V. Sapozhnikov.
\newblock {Skyrmion lattice in a magnetic film with spatially modulated
  material parameters}.
\newblock {\em Journal of Magnetism and Magnetic Materials}, 396:338--344,
  2015.

\bibitem{Sellmyer:2006}
D.J. Sellmyer, Y.~Xu, M.~Yan, Y.~Sui, J.~Zhou, and R.~Skomski.
\newblock Assembly of high-anisotropy {L10 FePt} nanocomposite films.
\newblock {\em Journal of Magnetism and Magnetic Materials}, 303:302--308,
  2006.

\bibitem{Stavrou:2016}
V.D. Stavrou, L.~N. Gergidis, A.~Markou, A.~Charalambopoulos, and
  I.~Panagiotopoulos.
\newblock {Micromagnetics of triangular thin film nanoelemets}.
\newblock {\em Journal of Magnetism and Magnetic Materials}, 401:716--723,
  2016.

\bibitem{Stier:2017}
Martin Stier, Wolfgang H\"ausler, Thore Posske, Gregor Gurski, and Michael
  Thorwart.
\newblock Skyrmion-anti-skyrmion pair creation by in-plane currents.
\newblock {\em Phys. Rev. Lett.}, 118:267203, Jun 2017.

\bibitem{Stosic:2017}
Dusan Stosic, Jeroen Mulkers, Bartel Van~Waeyenberge, Teresa~B. Ludermir, and
  Milorad~V. Milo\ifmmode \check{s}\else \v{s}\fi{}evi\ifmmode~\acute{c}\else
  \'{c}\fi{}.
\newblock Paths to collapse for isolated skyrmions in few-monolayer
  ferromagnetic films.
\newblock {\em Phys. Rev. B}, 95:214418, Jun 2017.

\bibitem{Parkin:2008}
S.~Stuart, P.Parkin, Masamitsu Hayashi, and Luc Thomas.
\newblock Magnetic domain-wall racetrack memory.
\newblock {\em Science}, 320(5873):190--194, 2008.

\bibitem{Tan:2016}
A.~Tan, J.~Li, A.~Scholl, E.~Arenholz, A.~T. Young, Q.~Li, C.~Hwang, and Z.~Q.
  Qiu.
\newblock Topology of spin meron pairs in coupled {Ni/Fe/Co/Cu(001)} disks.
\newblock {\em Physical Review B}, 94:014433, 2016.

\bibitem{Tomasello:2015}
R.~Tomasello, E.~Martinez, R.~Zivieri, L.~Torres, M.~Carpentieri, and
  G.~Finocchio.
\newblock A strategy for the design of skyrmion racetrack memories.
\newblock {\em Scientific Reports}, 4(6784):17137, 2014.

\bibitem{Wang:2012}
Y.~Wang, P.~Sharma, and A.~Makino.
\newblock Magnetization reversal in a preferred oriented {(111) L1 0 FePt}
  grown on a soft magnetic metallic glass for tilted magnetic recording.
\newblock {\em Journal of Physics Condensed Matter}, 24(7), 2012.

\bibitem{Xia:2018}
H.~Xia, C.~Song, C.~Jin, J.~Wang, J.~Wang, and Q.~Liu.
\newblock {Skyrmion motion driven by the gradient of voltage-controlled
  magnetic anisotropy}.
\newblock {\em Journal of Magnetism and Magnetic Materials}, 458:57--61, 2018.

\bibitem{Xia:2017}
Haiyan Xia, Chendong Jin, Chengkun Song, Jinshuai Wang, Jianbo Wang, and
  Qingfang Liu.
\newblock {Control and manipulation of antiferromagnetic skyrmions in
  racetrack}.
\newblock {\em J. Phys. D: Appl. Phys.}, 50:505005, 2017.

\bibitem{Yoo:2014}
J.W. Yoo, S.J. Lee, J.H. Moon, and K.J. Lee.
\newblock Phase diagram of a single skyrmion in magnetic nanowires.
\newblock {\em IEEE Transactions on Magnetics}, 50:1500504, 2014.

\bibitem{YuPnas:2012}
Xiuzhen Yu, Maxim Mostovoy, Yusuke Tokunaga, Weizhu Zhang, Koji 
  Matsui, Yoshio Kaneko, Naoto Nagaosa, and 
\newblock Magnetic stripes and skyrmions with helicity reversals.
\newblock {\em PNAS}, 109(23):8856--8860, 2012.

\bibitem{Zhang:2016}
S.~Zhang, A.~K. Petford-Long, and C.~Phatak.
\newblock {Creation of artificial skyrmions and antiskyrmions by anisotropy
  engineering}.
\newblock {\em Scientific Reports}, 6:131248, 2016.

\bibitem{Zhang:2015}
X.~Zhang, M.~Ezawa, and Y.~Zhou.
\newblock {Magnetic skyrmion logic gates: conversion, duplication and merging
  of skyrmions}.
\newblock {\em Scientific Reports}, 5:1500504, 2015.

\bibitem{Zhou:2015}
Y.~Zhou, E.~Iacocca, A.A. Awad, R.K. Dumas, F.C. Zhang, H.B. Braun, and
  J.~Akerman.
\newblock {Dynamically stabilized magnetic skyrmions}.
\newblock {\em Nature Communications}, 6:8193, 2015.

\end{thebibliography}

\end{document}